# Three-dimensional coordinates of individual atoms in materials revealed by electron tomography

Rui Xu[1†], Chien-Chun Chen[1†§], Li Wu[1†], M. C. Scott[1†], W. Theis[2†], Colin Ophus[3†],

Matthias Bartels[1], Yongsoo Yang[1], Hadi Ramezani-Dakhel[4], Michael R. Sawaya[5],

Hendrik Heinz[4], Laurence D. Marks[6], Peter Ercius[3] & Jianwei Miao[1*]

*[1]Department of Physics & Astronomy and California NanoSystems Institute, and University of California, Los Angeles, CA 90095, USA. [2]Nanoscale Physics Research Laboratory, School of Physics and Astronomy, University of Birmingham, Edgbaston, Birmingham B15 2TT, UK. [3]National Center for Electron Microscopy, Molecular Foundry, Lawrence Berkeley National Laboratory, Berkeley, California 94720, USA. [4]Department of Polymer Engineering, University of Akron, Akron, Ohio 44325, USA. [5]Howard Hughes Medical Institute, UCLA-DOE Institute of Genomics and Proteomics, Los Angeles, California 90095-1570, USA. [6]Department of Materials Science and Engineering, Northwestern University, Evanston, IL 60201, USA.*

*[†]These authors contributed equally to this work. [§]Present address: Department of Physics, National Sun Yat-sen University, Kaohsiung 80424, Taiwan. [*]Email: miao@physics.ucla.edu*

**Crystallography, the primary method for determining the three-dimensional (3D) atomic positions in crystals, has been fundamental to the development of many fields of science[1]. However, the atomic positions obtained from crystallography represent a global average of many unit cells in a crystal[1,2]. Here, we report, for the first time, the determination of the 3D coordinates of thousands of individual atoms and a point defect in a material by electron tomography with a precision of ~19 picometers, where the crystallinity of the material is not assumed. From the**

coordinates of these individual atoms, we measure the atomic displacement field and the full strain tensor with a 3D resolution of ~1nm$^3$ and a precision of ~10$^{-3}$, which are further verified by density functional theory calculations and molecular dynamics simulations. The ability to precisely localize the 3D coordinates of individual atoms in materials without assuming crystallinity is expected to find important applications in materials science, nanoscience, physics and chemistry.

In 1959, Richard Feynman challenged the electron microscopy community to locate the positions of individual atoms in substances[3]. Over the last 55 years, significant advances have been made in electron microscopy. With the development of aberration-corrected electron optics[4,5], scanning transmission electron microscopy (STEM) has reached sub-0.5 Å resolution in two dimensions[6]. In a combination of STEM[7-9] and a 3D image reconstruction method known as equal slope tomography (EST)[10,11], electron tomography has achieved 2.4 Å resolution and was applied to image the 3D core structure of edge and screw dislocations at atomic resolution[12,13]. More recently, transmission electron microscopy (TEM) has been used to determine the 3D atomic structure of gold nanoparticles by averaging 939 particles[14]. Notwithstanding these important developments, Feynman's 1959 challenge − 3D localization of the coordinates of individual atoms in a substance without using averaging or a priori knowledge of sample crystallinity − remains elusive. Here, we determine the 3D coordinates of 3,769 individual atoms in a tungsten needle sample with a precision of ~19 picometers and identify a point defect inside the sample in three dimensions. The acquisition of a high-quality tilt series with an aberration-corrected STEM and 3D EST reconstruction  allow us to trace individual atomic coordinates from the reconstructed intensity and refine the 3D atomic model. By comparing the coordinates of these individual atoms with an ideal body-centred-cubic (bcc) crystal lattice, we measure the atomic displacement field and the full strain tensor in three dimensions. Further experimental results, density functional theory (DFT) calculations and molecular dynamics (MD) simulations confirm that the displacement field and strain tensor are induced by a surface layer of tungsten carbide (WC$_X$) and the diffusion of carbon atoms several layers below the tungsten surface. While TEM, electron diffraction and holography can measure strain in nanostructures and devices with ≤1 nm resolution[15-17], they are mainly applicable in two dimensions. In order to image the 3D strain field, current methods, such as coherent diffractive imaging, compressive sensing electron tomography and through-focal annular dark-field imaging[18-20], use the phase in reciprocal space from crystalline samples[15,16]. Here, we directly image the 3D atomic positions and calculate the six element strain tensor in a material with a 3D resolution of ~1 nm$^3$ and a precision of ~10$^{-3}$, which are presently not attainable by any other methods.

The experiment was performed on an aberration corrected STEM operated in annular dark field (ADF) mode[21]. The sample was a tungsten needle, fabricated by electrochemical etching (Methods). By rotating the needle sample around the [011] direction from 0° to 180°, a tilt series of 62 angles was acquired with equal slope increments (Supplementary Fig. 1). The 0° (Fig. 1 inset) and 180° images of the tilt series are compared in Supplementary Fig. 2, indicating minimal change of the sample structure throughout the experiment. After correcting sample drift, scan distortion, and performing background subtraction on each image (Methods), the tilt series was aligned to a common rotation axis by a centre of mass method[12]. Only the apex of the needle (Fig. 1 inset and Supplementary Fig. 1) was used for the EST reconstruction due to the STEM depth of focus and to minimize dynamical scattering. Three different schemes were implemented to reconstruct our experimental data. First, a direct EST reconstruction was performed on the tilt series (termed the raw reconstruction). Second, 3D Wiener filtering was applied to the raw reconstruction to reduce the noise[22]. Third, the tilt series images were denoised by a sparsity based algorithm[23] (Supplementary Fig. 3) and then reconstructed by EST (Methods).

The EST reconstruction provides an estimate of the intensity distribution inside the tungsten tip, and further analysis known as atom tracing is needed to determine atomic coordinates. We traced and verified the 3D positions of individual atoms using two independent reconstructions: one using Wiener filtering and the other using sparsity denoising (Methods). During atom tracing, a 3D Gaussian function was fit to each local intensity maximum in both reconstructions. Then, we screened each of these plausible atoms by its fit to the average atom profile calculated from the corresponding reconstruction, yielding two sets of atom candidates (Methods). We selected only those in common between the two sets, totaling 3,641 atoms. For the non-common atom candidates, we evaluated the fit of each atom candidate with the profile of the average atom calculated from the raw reconstruction (Methods). An additional 128 atoms met our criteria, yielding a total of 3,769 traced atoms. After tracing the 3D positions of individual atoms, we implemented a refinement procedure to improve the agreement between the 3D atomic model and the raw experimental images (Methods). Each experimental image was transformed to obtain 62 Fourier slices, which were used to refine the 3D atomic model by iterating between real and reciprocal space[24] (Methods). Figure 1 and Supplementary Figs. 4-8 show the final refined 3D atomic model, consisting of 9 atomic layers along the [011] direction. The 3D profile of the average tungsten atom in the refined model is consistent with that of the average atom of the raw reconstruction (Supplementary Fig. 9).

To cross-check our procedure and evaluate the potential impact of dynamical scattering effects in ADF-STEM tomography[7,25], we performed multislice calculations[26] using the refined model for the same experimental conditions (Methods, Supplementary Fig. 10). By applying the exact reconstruction, atom tracing and refinement procedures, we obtained a new 3D atomic model from the 62 multislice calculated images, consisting of 3,767 atoms with only three misidentified atoms at the surface. Supplementary Fig. 11 shows a root-mean-square deviation (RMSD) of ~22 picometers between the experimental model and the new atomic model, suggesting that dynamical scattering has a minimal effect on our overall results within the measurement accuracy. We attribute the reduction of the dynamical scattering effects to the measurement of many images at different sample orientations (*i.e.* a rotational average) in our experiment (Supplementary Information).

Next, we estimated the precision of the 3D atomic positions determined from the experimental data. Based on the measured 0° image, we confirmed that the apex of the sample is strained (Supplementary Fig. 12) and selected the least strained region of the 3D atomic model. Using a cross-validation statistical method[27] to compare the atom positions in the selected region with the atomic sites of a best fit lattice, we determined a 3D precision of ~19 pm with contributions of approximately 10.5, 15.0 and 5.5 pm along the x-, y- and z-axes, respectively (Methods, Supplementary Fig. 13). The high-quality reconstruction and coordinates of individual atoms  both identify a point defect in the tungsten material in three dimensions. Figures 2a and b, show the reconstructed 3D intensity and surface renderings of three consecutive layers surrounding the point defect, located in layer 6. A quantitative comparison of the reconstructed intensity at the defect site and its surrounding atoms for all three reconstructions (raw, Wiener filtered and sparsity denoised) all strongly indicate a tungsten atom is not located at this site and it is not an error in the atom tracing. This is further confirmed by an unambiguous determination of the point defect in the 3D reconstruction of the multislice images, calculated from our experimental atomic model (Methods). While a substitutional point defect cluster of light atoms is energetically favourable in tungsten[28], a definitive identification of the substitutional atom species requires an experimental tilt series with a better signal to noise ratio.

Based on the 3D coordinates of the individual atoms, we measured the atomic displacement field of the sample (Methods). Figures 2c-e, Supplementary Fig. 14 and Movie 1 show the 3D atomic displacements calculated as the difference between the measured atomic positions and the corresponding ideal bcc lattice sites. The tip exhibits expansion along the $[0\bar{1}1]$ direction (x-axis) and compression along the $[100]$ direction (y-axis). The atomic displacements in the $[011]$ direction (z-axis) are less than half the magnitude of those along the x- and y-axes (Supplementary Movie 2). The 3D atomic displacements were used to determine the full strain tensor in the material. Calculation of the strain tensor requires differentiation of the displacement field making it more sensitive to noise. Therefore, we convolved the atomic displacement field with a 5.5-Å-wide 3D Gaussian kernel to increase the signal to noise ratio, but this also reduces the 3D spatial resolution to ~1nm. Figures 3a-d, shows the distribution of the atoms in layers 2-9 and the corresponding smoothed 3D displacement field along the x-, y- and z-axes, respectively. The six components ($\varepsilon_{xx}$, $\varepsilon_{yy}$, $\varepsilon_{zz}$, $\varepsilon_{xy}$, $\varepsilon_{xz}$ and $\varepsilon_{yz}$) of the full strain tensor (Figs. 3e-j) were determined from the smoothed displacement field with a precision of ~$10^{-3}$ (Supplementary Fig. 15) (Methods). The $\varepsilon_{xx}$, $\varepsilon_{yy}$, $\varepsilon_{xz}$ and $\varepsilon_{yz}$ maps exhibit features directly related to lattice plane bending, expansion along the x-axis and compression along the y-axis. Shear in the x-y plane is clearly visible in the $\varepsilon_{xy}$ map. Compared to the other components, the $\varepsilon_{zz}$ map is more homogeneous. By calculating the eigenvalues and eigenvectors using the full strain tensor, we obtained the principle strains to be approximately 0.81%, -0.87%, and -0.15% along the [0.074 0.775 -0.628], [0.997 -0.083 0.015], and [0.041 0.627 0.778] directions, respectively.

To understand the origin of the strain field, we projected the experimental 3D model along the [100], [0$\bar{1}$1] and [1$\bar{1}$1] directions. A comparison of the projected atomic positions with an ideal bcc lattice showed that the atomic displacements become larger closer to the surface (Supplementary Figs. 16a-c). This suggests that the tungsten positions and/or the chemical composition changed near the surface. Carbon was present on the tip and could have been intercalated in between the tungsten layers leading to a local expansion with octahedral coordination of the carbon, qualitatively in agreement with Supplementary Figs. 16d-i. To further explore this, we prepared another tungsten needle using the same sample preparation procedure except that carbon was deposited on the needle before heating to 1200°C in vacuum. ADF and bright-field STEM images along the [100] and [111] directions show bending of the atomic columns (Supplementary Fig. 17), and DFT calculations of surface tungsten carbide ($WC_x$) are in good agreement with the ADF images (Supplementary Text, Figs. 16d-i and Figs. 17a and c). Finally, we performed MD simulations of a tungsten needle with and without the presence of carbon (Supplementary Text and Fig. 18). The MD results show that the strain tensor approaches zero in the carbon-free needle (Supplementary Fig. 19). However, with intercalated carbon, the tungsten needle exhibits expansion and compression along the [0$\bar{1}$1] and [100] directions (Supplementary Fig. 20), respectively, in agreement with the experimental measurements (Fig. 3). Thus, our experimental results, MD simulations and DFT calculations all indicate that the strain in the tungsten needle is induced by surface $WC_x$ and the diffusion of carbon atoms several layers below the tungsten surface.

In conclusion, the 3D coordinates of thousands of individual atoms and a point defect in a material have been determined with a precision of ~19 picometers, where the crystallinity of the sample was not assumed. This allows direct measurements of the atomic displacement field and the full strain tensor with a 3D resolution of ~1 nm and a precision of $10^{-3}$, which were further verified by DFT calculations and MD simulations. Although a tungsten needle sample was used here as a proof-of-principle, our method can be applied to a wide range of materials that can be processed into small volumes, including nanoparticles, nanowires, nanorods, thin films, and needle-shaped specimens used in atom probe tomography[29]. While we resolved the positions of individual tungsten atoms in this experiment, numerical simulation results indicate that this method can also be used to determine the 3D coordinates of individual atoms in amorphous materials[30]. The ability to precisely localize the 3D coordinates of individual atoms in materials without assuming crystallinity, identify point defects in three dimensions, and measure the 3D atomic displacement field and the full strain tensor, coupled with DFT calculations and MD simulations, is expected to transform our understanding of materials properties and functionality at the most fundamental scale.

**Acknowledgements** We thank U. Dahmen, J. Du, L Deng, E. J. Kirkland, R. F. Bruinsma and L. A. Vese for stimulating discussions. This work was primarily supported by the Office of Basic Energy Sciences of the U.S. Department of Energy (Grant No. DE-FG02-13ER46943). This work was partially supported by NSF (DMR-1437263 and DMR-0955071) as well as ONR MURI (N00014-14-1-0675). L.D.M acknowledges support from the DOE (Grant No. DE-FG02-01ER45945). ADF-STEM imaging was performed on TEAM I at the Molecular Foundry, which is supported by the Office of Science, Office of Basic Energy Sciences of the U.S. Department of Energy under Contract No. DE-AC02—05CH11231. H.R. and H.H. acknowledge the allocation of computing resources at the Ohio Supercomputer Center.

## Figure legends

**Figure 1**. **3D positions of individual atoms in a tungsten needle sample revealed by electron tomography.** The experiment was conducted using an aberration-corrected STEM. A tilt series of 62 projections was acquired from the sample by rotating it around the [011] axis. The inset shows a representative projection at 0˚. After post-processing, the apex of the sample (labelled with a rectangle in the inset) was reconstructed by the EST method. The 3D positions of individual atoms were then traced from the reconstructions and refined using the 62 experimental projections. The 3D atomic model of the sample consists of 9 atomic layers along the [011] direction, labelled with crimson (dark red), red, orange, yellow, green, cyan, blue, magenta and purple from layers 1 to 9, respectively.

**Figure 2**. **3D determination of a point defect and the atomic displacements in the tungsten needle sample. a** and **b**, 3D density and surface renderings of a point defect in the tungsten sample (diamond-shaped region in (**c**)), clearly indicating no tungsten atom density at the defect site. **d**, **e** and **f**, 3D atomic displacements in layer 6 of the tungsten sample along the x-, y- and z-axes, respectively, exhibiting expansion along the $[0\,\overline{1}\,1]$ direction (x-axis) and compression along the [100] direction (y-axis). The atomic displacements in the [011] direction (z-axis) are smaller than those in the x- and y-axes. The atoms with white dots are excluded for displacement measurements due to their relatively large deviations from a bcc lattice (Methods). The 3D atomic displacements in other layers are shown in Supplementary Fig. 13.

**Figure 3**. **3D strain tensor measurements in the tungsten needle sample. a**, Atoms in layers 2-9 used to determine the 3D strain tensor, where layer 1 and other surface atoms in red are excluded for displacement field and strain measurements. **b**, **c** and **d**, 3D lattice displacement field for layers 2-9 along the x-, y- and z-axes, respectively, obtained by convolving the 3D atomic displacements with a 5.5-Å-wide 3D Gaussian kernel to reduce the noise and increase the precision. Expansion along the $[0\bar{1}1]$ direction (x-axis) and compression along the [100] direction (y-axis) are clearly visible. **e**, **f**, **g**, **h**, **i** and **j**, Maps of the six components of the full strain tensor, where $\varepsilon_{xx}$, $\varepsilon_{yy}$, $\varepsilon_{xz}$ and $\varepsilon_{yz}$ exhibit features directly related to lattice plane bending, expansion and compression along the along the x- and y-axis, respectively. $\varepsilon_{xy}$ shows shear in the x-y plane and $\varepsilon_{zz}$ is more homogenously distributed.

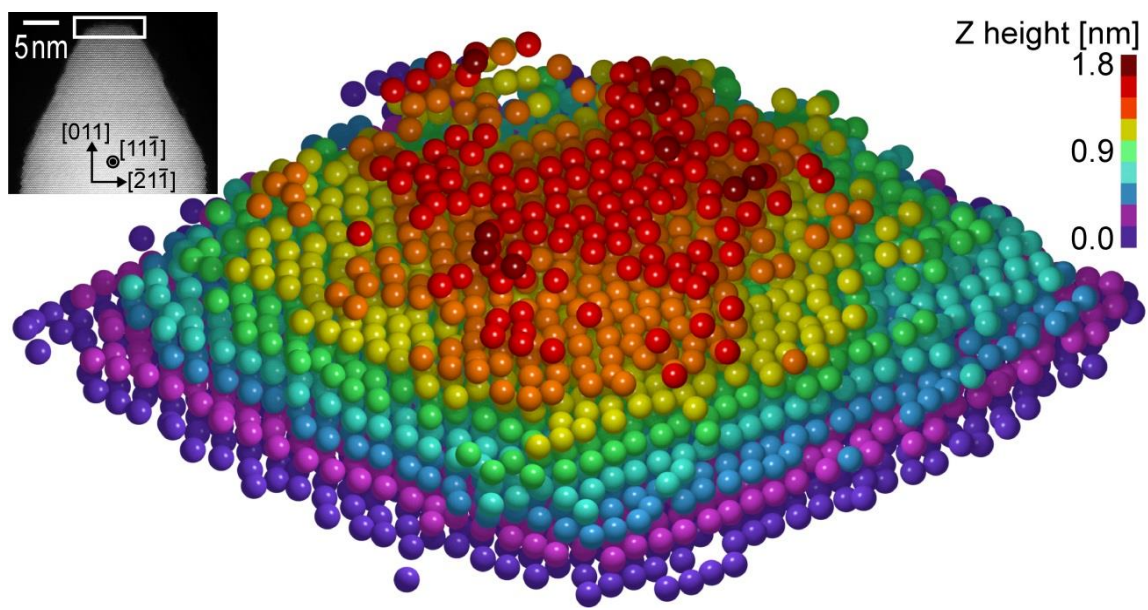

**Figure 1**

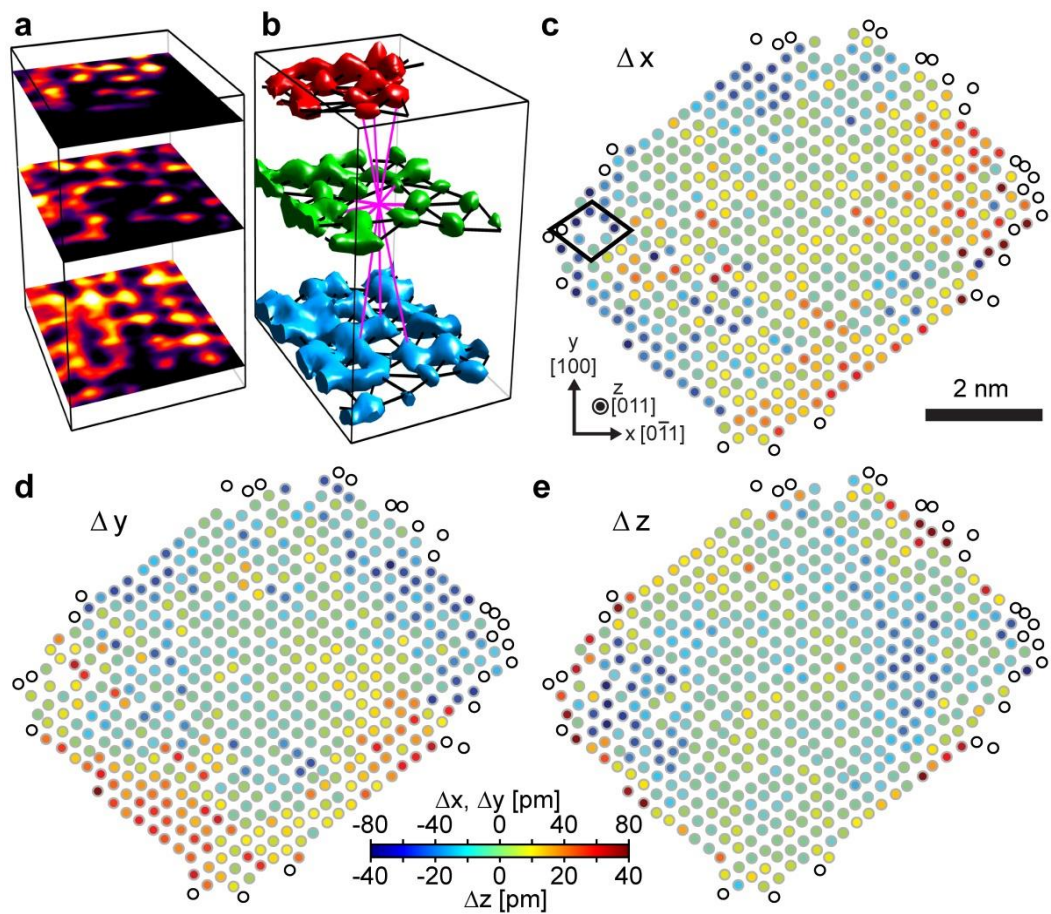

**Figure 2**

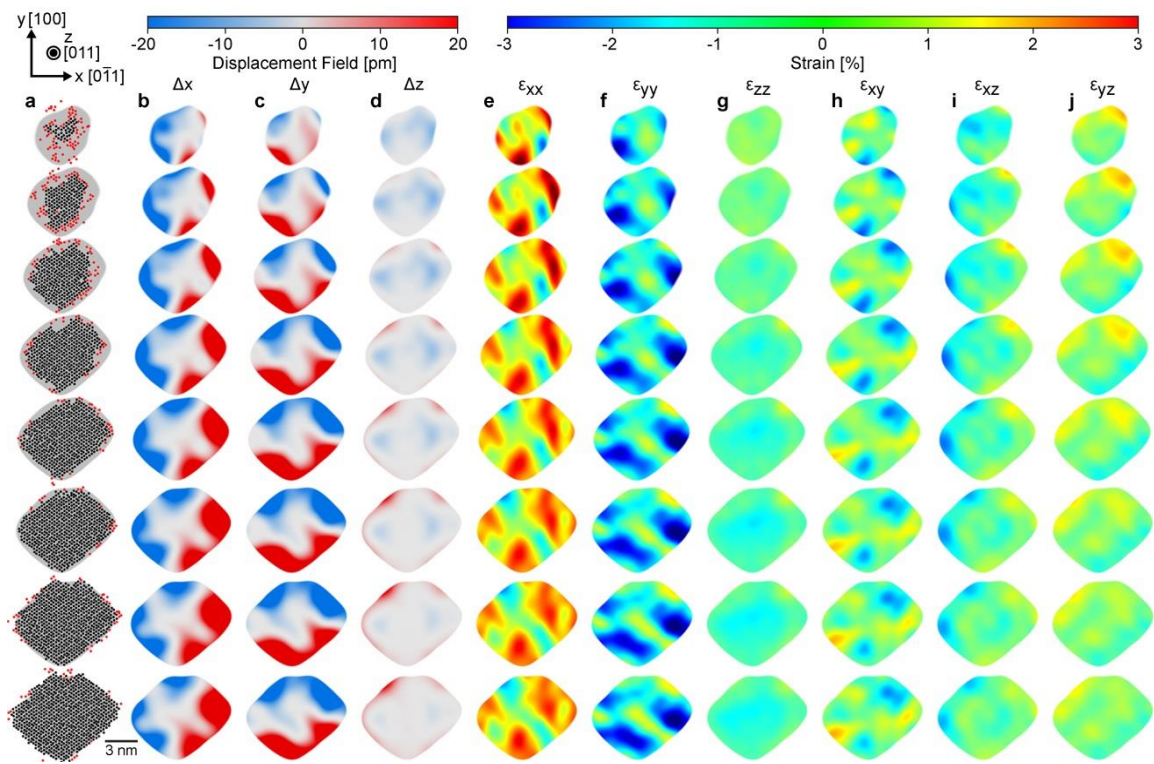

**Figure 3**

## METHODS

**Sample Preparation**. A tungsten wire with 99.95% purity was annealed under tension until melted, creating a large crystalline domain with [011] preferentially aligned along the wire axis. The 250 µm wire was then electrochemically etched in a NaOH solution using a dedicated etching station with an electronic cutoff circuit to form a sharp tip with a <10nm diameter. The wire was then plasma cleaned in an Ar/O$_2$ gas mixture and heated to 1000 °C in vacuum (~$10^{-5}$ Pa) to remove the oxide layer generated by the plasma cleaning. The wire was mounted in a 1 mm sample puck appropriate for the TEAM stage.

**Data Acquisition**. Tomographic data was acquired using the TEAM I at the National Center for Electron Microscopy in The Molecular Foundry operated at 300 kV in ADF-STEM mode (convergence semi-angle: 30 mrad; detector inner and outer semi-angles: 38 mrad and 200 mrad; aberration-corrected probe size: ~50 pm; beam current: 70 pA). The TEAM stage is a tilt-rotate design with full 360° rotation about both axes. The tomography rotation axis was chosen to be the [011] crystallographic axis of the tungsten sample. A tomographic tilt series of 62 images was acquired from the tungsten needle sample at EST angles, covering the complete angular range of ±90°. Two images of 1024x1024 pixels each with 6 μs dwell time and 0.405 Å pixel resolution were acquired at each angle. To reduce the radiation dose, a low-exposure acquisition scheme was implemented[12]. When focusing an image, a nearby sample was first viewed, thus reducing unnecessary radiation dose to the sample under study. The total dose used in the tungsten needle data set was comparable to that reported before[12,13]. To monitor the consistency of the tilt series, we measured the 0° images of the tungsten sample before and after the acquisition of the full data set, showing the consistency of the sample structure throughout the experiment (Supplementary Fig. 2).

**ADF-STEM Image Preprocessing**. Preprocessing of images involved compensating for constant sample drift and STEM scan distortions. Sample drift was determined from the relative shift of the pairs of images taken for each EST angle. The STEM scan distortion was determined from the Fourier transform of a region 18.5 nm from the apex in the [$11\bar{1}$] image assuming a bulk bcc tungsten lattice structure in this region. The resulting linear mapping required to correct for the measured drift and to achieve square pixels of 0.405 Å pixel size was decomposed into a product of shear transformations and pure x and y axis scaling operations which were applied to the ADF-STEM images using Fourier methods for shear[31] and scaling operations. Due to the nature of the 2-axis TEAM stage design, the tomography axis has a different in-plane orientation in the ADF-STEM image for each EST angle. The Fourier transform of a region 12.5 nm from the apex in the individual images was used to determine the orientation of the [011] tomography axis. The images were individually rotated using Fourier methods[31] to align the [011] direction along the image vertical.

**Background Subtraction and Denoising of Individual Images**. To estimate the background and noise level in each experimental image, we adopted a noise model for each pixel, $Y = \alpha P(n_e) + N(\mu_b, \sigma_b)$, where $Y$ is the intensity counts, $\alpha$ the counts per electron, $P(n_e)$ the Poisson distribution of $n_e$ electrons, and $N(\mu_b, \sigma_b)$ the normal distribution of the background with a mean ($\mu_b$) and standard deviation ($\sigma_b$). To verify this noise model, we acquired 126 images of a sample for the same experimental conditions with TEAM I. Using the 126 images, we calculated $P(n_e)$ for various pixels and confirmed that $P(n_e)$ was a Poisson distribution. Next, we applied this noise model to each experimental image to estimate the background and the corresponding $n_e$. After performing background subtraction for each image, we obtained 62 images which would be used for the raw EST reconstruction and further denoising. Our denoising process was implemented by first transforming Poisson noise to Gaussian noise[32] and then applying a sparsity based algorithm that has been widely used in the image processing field[23]. Supplementary Fig. 3 shows the 0° image before and after denoising as well as their difference, indicating that the denoising process did not introduce any visible artifacts.

**EST Reconstructions**. The 62 raw and denoised images were reconstructed by EST with the following procedure. i) The 62 images were projected to the tilt axis to generate a set of 1D curves, which was aligned by cross-correlation with 0.1 pixel steps. ii) The images were then projected to an axis perpendicular to the tilt axis to produce another set of 1D curves, which was aligned by a CM method with 0.1 pixel per step[12]. Steps i) and ii) were repeated until no further improvement could be made. iii) The aligned images were reconstructed by EST with positivity as a constraint and 500 iterations. We found that the reconstruction was slightly improved by not enforcing a support constraint. iv) The 3D reconstruction was projected back to calculate the corresponding 62 images. The calculated images were used as references to further align the experimental images. Steps iii) and iv) were repeated until there was no further improvement. v) A loose support (i.e. a boundary slightly larger than the true envelop of the sample) was estimated from the final reconstruction and the intensity outside the loose support was removed.

**Tracing of 3D Atom Positions**. The 3D positions of individual atoms were traced using a two-step approach. In step 1, we first identified common atoms in two independent 3D structures. Structure one was reconstructed from 62 denoised images and structure two was obtained by taking the square root of the product of the raw reconstruction and the Wiener filtered reconstruction ($\lambda$=1)[22]. Step 1 consists of the following sub-steps. i) The positions of all local maxima in each 3D reconstruction were identified and sorted from the highest to lowest intensity. ii) Starting from the highest intensity, a 3D Gaussian function was fit to the local maximum. If a minimum distance constraint (the distance of two neighboring atoms $\geq$ 2Å) was satisfied, the peak of the fit to a Gaussian function was chosen as a plausible atom position and the Gaussian function was then subtracted from the corresponding reconstruction. Sub-steps i) and ii) were repeated until two complete sets of plausible atom positions were obtained from two independent reconstructions. iii) Next, the average atom profile was generated by summing up a large number of plausible atoms for each reconstruction, omitting extraordinarily high and low peaks. A 3D Gaussian was then fit to the average atom profile. iv) Every plausible atom in each complete set was checked with the 3D Gaussian function of the average atom,

$$R_{atom} = \frac{\sum_{\vec{r}}\left| f(\vec{r}) - b_{ave} \right|}{\sum_{\vec{r}}\left| f(\vec{r}) - f_{ave}(\vec{r}) \right|} \ , \qquad (1)$$

where $f(\vec{r})$ represents a Gaussian approximation to the shape of a plausible atom, $f_{ave}(\vec{r})$ a Gaussian approximation of the average atom for a corresponding reconstruction, and $b_{ave}$ the background of the Gaussian function fit to the average atom. If $R_{atom} \geq 1$ (indicating the candidate atom is closer to the average atom than to the background), the plausible atom was selected as an atom candidate. v) The atom candidates in the two datasets were quantitatively compared to each other. The common pairs of atoms in the two datasets with deviations smaller than the radius of the tungsten atom (1.39Å) were selected as atoms. The position of each selected atom was determined by averaging the common pair of atomic positions. vi) Sub-steps i-v) were repeated until there was no further improvement and 3,641 common atoms were identified.

After finding the common atoms, we examined the non-common atoms in two independent datasets in step 2, which consists of the following sub-steps. i) The non-common atoms in the two atom candidate datasets were identified. ii) The average atom profile was obtained from the raw reconstruction, to which a 3D Gaussian function was fit. Eq. (1) was used to examine the non-common atoms. Those with $R_{atom} \geq 1$ and also satisfying the minimal distance constraint were chosen as atoms. iii) We checked each of the chosen atoms with both the raw reconstruction and the reconstruction obtained from denoised images, and removed false atoms. iv) Sub-steps i-iii) were repeated until no further improvement could be made, resulting in an additional 128 atoms being identified. Finally, combining steps 1 and 2, we obtained a traced 3D atomic model with a total of 3,769 atoms.

**3D Atomic Model Refinement.** The traced atomic model was refined by using the following steps. i) The 62 raw experimental images were converted to Fourier slices, $F_{obs}^n(\vec{q})$ with $n = 1,..., 62$, by a fast Fourier transform. ii) The corresponding 62 Fourier slices were calculated from the traced atomic model by

$$F_{calc}^n(\vec{q}) = \sum_{j=1}^{M} f_e(q) e^{-B'q^2/4 - 2\pi i \vec{r}_j \cdot \vec{q}} \ , \qquad (2)$$

where $F_{calc}^n(\vec{q})$ represents the $n^{th}$ calculated image, $M = 3769$ is the number of atoms, $f_e(q)$ the electron scattering (form) factor of tungsten[26], $\vec{r}_j$ the position of the $j^{th}$ atom, and $B'$ accounts for the thermal motion of the atom, the electron probe size (50 pm) and the reconstruction error. We note that within a tight-binding expansion, the leading term in the scattering is almost the same as the kinematical potential, as can be seen by comparing the limits for small thicknesses (see also the later discussion on dynamical effects). Since our model consists of one type of atom, every atom was treated as isotropic and identical.

iii) The experimental and calculated Fourier slices were quantitatively compared by the functional

$$E = \sum_{n=1}^{62} \sum_{\vec{q}} | F_{obs}^n(\vec{q}) - F_{calc}^n(\vec{q}) |^2 . \qquad (3)$$

which was minimized with respect to the atomic position ($\vec{r}_j$) by a gradient descent method. iv) The total potential energy $U$ of the system in an embedded atom model[33,34] was used as a regularization to independently monitor the refinement

$$U = \frac{1}{2} \sum_{i,j,i \neq j} \phi_{ij}(r_{ij}) + \sum_i J_i(\rho_i) \qquad (4)$$

where $\phi_{ij}$ represents the pair energy between atoms $i$ and $j$ separated by $r_{ij}$, and $J_i$ the embedding energy for an atom $i$ in a site with electron density $\rho_i$. The parameters for calculating $\phi_{ij}$ and $J_i$ for tungsten atoms were obtained elsewhere[33]. The potential energy form of Eq. (4) has been widely used in MD simulations, known as the embedded atom method[34]. The total potential $U$ was not used as a constraint in our refinement, but was recorded for monitor purposes. The sum of the experimental and potential energy terms ($E$ and $U$) was used to optimize the number of iterations. iv) After obtaining a refined atomic model, we compared it with the independent reconstructions, manually adjusted the positions of <1% of the atoms, and obtained an updated model. The manual adjustment of a very small percentage of atoms has been widely used in the refinement in protein crystallography[24]. The updated atomic model was refined once more with the raw experimental images. This step was repeated until the average $R_1$ factor could not be further reduced (Supplementary Table 1). $B'$ in our final refinement was 17.2 Å$^2$. The relatively large value of $B'$ is due to the convolution with the electron probe size, thermal vibrations and the reconstruction error.

**Multislice STEM Calculations**. We performed multislice simulations based on the refined atomic model[26]. The atomic model was placed in a rectangular prism super cell (a=110.0Å; b=27.50Å; c=110.0 Å). The super cell was divided into multiple slices with different atomic positions along the z-axis each 1.6 Å thick and the x-y plane was discretized into 2048×512 pixels. The experimental parameters were used in the multislice simulations (electron energy: 300keV; $C_3$: 0mm; $C_5$: 5mm; convergence semi-angle: 30 mrad; detector inner and outer semi-angles: 38 mrad and 200 mrad). The electron beam propagated along the z-axis and each ADF-STEM image was generated by a raster-scan of 271×61 pixels in the x-y plane with 0.405Å per pixel. By rotating the atomic structure along the y-axis, a tilt series of images was computed for the experimental tilt angles. To simulate realistic experimental conditions, a tilt angle offset was continuously changed from 0° to 0.5° during the calculation of the tilt series. For each tilt angle, we employed the frozen phonon approach and averaged 20 phonon configurations to obtain a multislice image. The multislice image was convolved with a 3x3 pixel Gaussian function to account for the electron probe size, thermal vibrations, and other incoherent effects making the contrast in the simulation comparable to the experimental one. Poisson noise was added based on the experimental electron dose. Following this procedure, a tilt series of 62 ADF-STEM images was obtained. Supplementary Fig. 10 shows the experimental and multislice images at 0°. By using the same reconstruction, atom tracing and refinement methods, we obtained a new 3D atomic model from the 62 multislice images, in which only three atoms were misidentified at the surface. Supplementary Fig. 11 shows a histogram of the atomic deviation between the original and new atomic models, indicating a RMSD of ~22 picometers.

**Precision Estimation of Atomic Displacement Measurements**. In the flattest region of the sample where the lattice was closest to the ideal bcc, we estimated the displacement precision as the RMSD of our measured atomic positions from the site positions of a best-fit lattice. To determine which region of the sample was closest to an ideal bcc lattice, we used a cross-validation (CV) procedure[27]. In this procedure, a subset of the atomic positions was first selected for testing by determining all sites within a given fitting radius. We then calculated a best-fit lattice using a randomly selected set of half of these sites, and used it to predict the location of the remaining half. The CV score is equal to the RMSD of these predicted sites from the corresponding measurements. This procedure was repeated thousands of times using a new randomly generated half subset each time, to determine the mean CV score. This procedure was then repeated for various different fitting radii or equivalently the number of sites included. The purpose of a CV examination is to determine how many sites should be included in a linear best-fit lattice such that the lattice is neither under-fit (too few fitting parameters relative to the number of measurements) or over-fit (too many fitting parameters). When this condition is met, the CV score reaches a minimum. The depth of the minimum roughly indicates how close to an ideal lattice the measurement is. Supplementary Fig. 13 shows the CV score reaches a minimum when 23 sites are included in the lattice fitting. An upper bound for the precision can be estimated using the RMSD fitting error when 23 sites are included (Supplementary Fig. 13). For our experimental dataset, this precision was ~19 pm. This estimate can be further broken down into the three precision values: approximately 10.5, 15 and 5.5 pm along the x-, y- and z-axes, respectively. These values represent an upper estimate for the precision because no part of the tip forms an ideal bcc lattice. The smaller error along the z-axis is because the x-y plane contains information from only 62 images, but the z-axis has no missing information. The slight difference between the precision estimate (~19 pm) from the experimental data and the RMSD (~22 pm) obtained from multislice simulations is because i) in our multislice simulations, the tilt angle offset was continuously changed from 0° to 0.5° during the calculation of the tilt series. This offset is slightly larger than our experimental precision (<0.2°); and ii) a Gaussian function was used to convolve with each of the 62 multislice images (Supplementary Fig. 10). Our numerical simulations indicate that if we decrease the width of the Gaussian function, the RMSD can be reduced.

**3D Determination of the Strain Tensor from Measured Atomic Positions**. The strain present in the reconstruction was measured using the following procedure. First, an ideal bcc lattice was estimated by a least squares fit of the atom positions near the tip center. Then, each atom's displacement from its nearest-neighbours (up to 8) was calculated. Atoms that fell within one quarter of the nearest-neighbour bond length (0.69 Å) relative to the fitted bcc lattice vectors were marked as belonging to the bcc lattice. The lattice was then refit using a least squares method. These two steps were repeated until a self-consistent identification of the bcc lattice was obtained, which included 90.42% of the atomic positions (3,408 out of 3,769 sites). All atoms not included in the bcc lattice fit were located at the tip surface.

Next, the atomic displacements were calculated as the difference between the measured atomic positions and the corresponding ideal bcc lattice sites (Supplementary Fig. 14). The displacements were then interpolated onto a cubic grid using kernel density estimation[35]. In this method, the atomic intensity was first estimated by calculating a weighted sum at each voxel of a 3D Gaussian distribution, with a standard deviation equal to the kernel width. Then, each of the displacement field ($\Delta x, \Delta y$ and $\Delta z$) was estimated for each voxel by a weighted sum of a 3D Gaussian distribution multiplied by each displacement measurement, divided by the intensity estimate. For example, a set of $N$ points at distances $x_i$ with displacements of $d_i$ along the x-axis, would have a displacement field estimate $\Delta x$ of

$$\Delta x = \frac{\sum_{i=1}^{N} d_i \exp\left(-\dfrac{x_i^2}{2\sigma^2}\right)}{\sum_{i=1}^{N} \exp\left(-\dfrac{x_i^2}{2\sigma^2}\right)}. \qquad (5)$$

To produce a smooth estimate of the displacement field (a requirement for differentiation), a 3D Gaussian kernel with $\sigma = 5.5$ Å was chosen. The use of a Gaussian kernel increases the signal to noise ratio and precision, but reduces the resolution. Resolution is roughly twice of the kernel bandwidth and was ~1 nm in our measurements. Finally, the 3D strain tensor was calculated by numerical differentiation of the displacement field (Fig. 3), where the edge of the experimental displacement and strain fields were masked at approximately one third of the intensity value at the center of the tip.

**Precision Estimation of the Strain Tensor Measurements**. To estimate the strain measurement precision, we used numerical analysis and Monte Carlo simulations in one, two and three dimensions. These results are shown in Supplementary Fig. 15 for evenly spaced measurements along a line, a square lattice and a cubic lattice. By defining the relative kernel size ($k$) in terms of the lattice spacing ($a$), $k = \sigma/a$, we determined the dependence of the ratio between the RMSD ($\delta_{disp}$) and the strain measurement precision ($\delta_{strain}$) times the lattice spacing on the relative kernel size. The result is a simple power law for all three dimensions. The given numerical prefactors are close approximations. Since our measurements were performed on a bcc lattice, the atomic intensity is $\sqrt[3]{2}$ times that of a simple cubic lattice. Therefore, in three dimensions, the strain measurement precision is approximately

$$\delta_{strain} = \frac{\delta_{disp}}{10 \sqrt[3]{2}\, a\, k^{2.5}}. \qquad (6)$$

Our best fit lattice has side length $a = 3.18$ Å, and we used a kernel width equal to twice the nearest-neighbour distance ($k = \sigma/a = 1.73$). These values yield a strain measurement precision of

$$\delta_{strain} = \frac{0.19\ \text{Å}}{10 \sqrt[3]{2}\, (3.18\ \text{Å})\, (1.73)^{2.5}} = 0.12\% . \qquad (7)$$

The kernel size was chosen to keep the strain measurement precision well-below the measured peak values at the expense of reduced resolution. For example, halving the kernel size to a single nearest-neighbour length would change the strain measurement precision to 0.68%.